\documentclass[11pt]{article}
\usepackage[letterpaper,margin=1in]{geometry}

\usepackage{amsmath}
\usepackage{amssymb}
\usepackage{graphicx}
\usepackage{amsthm}
\usepackage{booktabs}
\usepackage{enumitem}
\usepackage{xcolor}
\usepackage{tikz}
\usepackage{eso-pic}
\usepackage{url}

\providecommand{\citenamefont}[1]{#1}

\newtheorem{theorem}{Theorem}[section]

\newtheorem{proposition}[theorem]{Proposition}

\newtheorem{definition}[theorem]{Definition}

\newcommand{\fito}{\textsc{fito}}
\newcommand{\oae}{\textsc{oae}}
\newcommand{\rdma}{\textsc{rdma}}
\newcommand{\lww}{\textsc{lww}}

\newcommand{\crdt}{\textsc{crdt}}

\title{\textbf{The Semantic Arrow of Time, Part~IV:}\\[0.3em]{\large Why Transactions Fail}}
\author{Paul Borrill \\ D\AE D\AE LUS}
\date{02026-FEB-27}

\usepackage{natbib}
\usepackage[hidelinks]{hyperref}

\begin{document}
\maketitle

\begin{center}
\large\itshape File Synchronization, Email, and the Fragility of Memory
\end{center}
\vspace{1em}

\begin{abstract}
\noindent
This is the fourth of five papers comprising \emph{The Semantic Arrow
of Time}.  Parts~I--III established that computing's hidden arrow of
time is semantic rather than thermodynamic~\citep{borrill2026-partI},
that bilateral transaction protocols create causal order through
a mandatory reflecting phase~\citep{borrill2026-partII}, and that
\rdma{}'s completion semantics implement the \fito{} category mistake
at industrial scale~\citep{borrill2026-partIII}.

This paper traces the consequences of the \fito{} category mistake
beyond the data center, into the systems that people use every day.
We examine three domains where forward-only temporal assumptions
destroy meaning: \emph{file synchronization}, where cloud platforms
silently delete user content because last-writer-wins cannot represent
distributed causality; \emph{email}, where timestamp-based ordering
produces phantom messages, causality violations, and stuck
synchronization; and \emph{memory}---both human and artificial---where
reconstructive processes that operate without transactional guarantees
produce systematic semantic corruption.

In each domain, we identify the same structural pattern: a system
that commits state changes forward in time without a reflecting phase,
and that therefore cannot distinguish between successful semantic
integration and mere temporal succession.  The pattern is not
coincidental.  It is the \fito{} category mistake operating at
different scales: bytes in a NIC buffer, files in a cloud, messages
in an inbox, engrams in a hippocampus, tokens in a transformer.

We conclude that the semantic arrow of time is violated whenever a
system treats the forward flow of information as sufficient evidence
of meaning.  Part~V will show how the Leibniz Bridge provides a
unified framework for closing this gap across all five domains.
\end{abstract}

% ------------------------------------------------------------------
\section[Introduction]{Introduction: When Meaning Is Lost}
\label{sec:intro}
% ------------------------------------------------------------------

The previous papers in this series examined the semantic arrow of
time at increasing levels of engineering specificity: philosophical
foundations (Part~I), link-level protocol design (Part~II), and
interconnect hardware (Part~III).  At each level, the same pattern
emerged: systems that assume forward temporal flow is sufficient for
establishing meaning produce systematic semantic corruption.

This paper changes perspective.  Instead of examining the engineering
substrate, we examine the \emph{consequences}---what happens to the
people, data, and organizations that depend on systems built on
\fito{} assumptions.  We present three case studies, each operating
at a different scale but exhibiting the same structural pathology.

\emph{File synchronization} is the most immediate example.  Every
major cloud platform---iCloud, Google Drive, Dropbox, OneDrive---uses
some variant of last-writer-wins (\lww{}) to resolve conflicts.  As
the companion paper \emph{Why iCloud Fails}~\citep{borrill2026-icloud}
demonstrates in detail, \lww{} is the \fito{} category mistake applied
to files: it projects a distributed causal graph onto a linear temporal
chain and treats the projection as authoritative.  The result is silent
data destruction.

\emph{Email} operates at a larger temporal scale.  Users compose,
read, archive, and delete messages across multiple devices over hours
and days.  Every email system relies on timestamps to order these
operations~\citep{borrill2026-email}.  When clocks disagree---as they
always do in distributed systems---the result is phantom messages,
lost read states, and causal inversions: replies that appear before
the messages they respond to.

\emph{Memory}---both human and artificial---operates at the largest
scale.  Human memory is reconstructive, not reproductive: we do not
replay recordings; we reassemble fragments from distributed neural
ensembles~\citep{bartlett1932}.  This reconstruction operates without
transactional guarantees, producing the false memories, confabulations,
and temporal distortions that Schacter~\citep{schacter2001} catalogued
as the ``seven sins of memory.''  Large language models exhibit an
isomorphic pathology: autoregressive token generation is inherently
\fito{}, producing hallucinations that are structurally identical to
the confabulations of a Korsakoff patient filling gaps in episodic
memory.

The remainder of this paper examines each domain in turn, identifies
the common \fito{} pattern, and shows how the \oae{} framework from
Part~II provides the structural vocabulary for understanding---and
eventually correcting---the failures.

% ------------------------------------------------------------------
\section[File Synchronization]{File Synchronization: The Illusion of Seamlessness}
\label{sec:filesync}
% ------------------------------------------------------------------

Cloud file synchronization promises a simple abstraction: your files
are everywhere, always up to date, always consistent.  The promise is
a lie---not because the engineering is poor, but because the
abstraction is structurally impossible under \fito{} assumptions.

\subsection{The Structural Thesis}

The companion paper \emph{Why iCloud Fails}~\citep{borrill2026-icloud}
establishes a precise structural thesis:

\begin{quote}
\emph{Projecting a distributed causal graph onto a linear temporal
chain destroys essential structure.  Information is lost.  The losses
manifest as corruption, conflicts, stalls, and silent data
destruction.}
\end{quote}

This projection is not unique to iCloud.  Every cloud sync platform
that uses timestamp-based conflict resolution performs the same
projection.  The differences between platforms are differences of
engineering quality within a shared architectural error.%

\subsection{Last-Writer-Wins as \fito{}}

\lww{} conflict resolution is \fito{} applied to files.  When two
devices modify the same file, the modification with the later
timestamp wins.  This strategy has three structural defects:

\begin{enumerate}[leftmargin=1.2cm]
  \item \textbf{Clock dependence.}  \lww{} assumes globally
    synchronized clocks.  Consumer devices routinely exhibit clock
    skew of several seconds; virtualized environments can experience
    jumps of minutes or hours during migrations or snapshots.  The
    ``later'' write may not be later in physical time.
  \item \textbf{Semantic blindness.}  \lww{} does not examine
    \emph{what} was modified.  If Alice edits paragraph~3 and Bob
    edits paragraph~7, both edits are valid and should be merged.
    \lww{} picks one and discards the other.
  \item \textbf{Silent loss.}  The discarded version vanishes without
    notification.  As Dominik Mayer documented~\citep{mayer2023},
    iCloud Drive silently deletes user content when version conflicts
    arise.  The ``core issue is the product decision to not be
    transparent about version conflicts.''\footnote{During preparation
    of this paper the authors observed a real-time \lww{} conflict:
    two copies of a PDF with identical timestamps, one from a VM write
    and one from the iCloud sync daemon.  See Incident~8 in
    Appendix~A of~\citep{borrill2026-icloud}.}
\end{enumerate}

In the \oae{} framework of Part~II, \lww{} is a protocol that signals
completion at $T_4$---the file has been placed in cloud storage---without
ever reaching $T_6$---semantic agreement between all devices about what
the file \emph{means}.  The reflecting phase is absent.  The user's
``Saved'' indicator is a completion signal masquerading as a commitment
signal, precisely as in the \rdma{} completion fallacy of Part~III.

\subsection{The Five Incompatibilities}

\emph{Why iCloud Fails} identifies five interlocking incompatibilities,
each rooted in the same \fito{} structural error:

\paragraph{Time Machine.}
Time Machine archives filesystem trees indexed by wall-clock time.
Cloud sync negotiates distributed event graphs.  A tree snapshot is
not equivalent to a distributed agreement state.  Restoring from
Time Machine while iCloud is active can cause the sync daemon to
treat cloud state as authoritative, destroying the ``correct past''
that Time Machine preserved.%

\paragraph{Git.}
Git depends on POSIX filesystem semantics---atomic lock files via
\texttt{open(O\_CREAT|O\_EXCL)}, stable paths, and the write-sync-rename
pattern.  iCloud treats lock files as ordinary files and syncs them,
propagates intermediate states that Git intends to be invisible, and
resolves parallel modifications by creating numbered suffixes
(\texttt{main~2}) that Git never expects.

\paragraph{Automated toolchains.}
Build systems assume that file existence implies file presence.
iCloud's dataless files---metadata present, data extents absent---break
this assumption silently.  A dataless file returns valid metadata from
\texttt{stat()} but triggers a network fetch when read.

\paragraph{Developer workflows.}
Any workflow that assumes POSIX atomicity, advisory locking, or
consistent directory listings is incompatible with cloud sync's
eventual-consistency model.

\paragraph{The 366\,GB archive.}
The companion paper documents a specific case: 366\,GB of divergent
state accumulated through normal use, representing the accumulated
damage of years of \fito{}-based conflict resolution operating on a
large corpus of documents.

\subsection{What Would Fix It}

The structural fix is not better engineering within the \lww{}
framework.  It is the replacement of \lww{} with a protocol that
includes a reflecting phase:

\begin{enumerate}[leftmargin=1.2cm]
  \item \textbf{Vector clocks} or \textbf{hybrid logical clocks}
    to track causal dependencies rather than wall-clock order.
  \item \textbf{Conflict materialization}: when files diverge, present
    both versions to the user rather than silently choosing one.
  \item \textbf{Semantic merge}: for structured documents, merge at
    the semantic level (paragraphs, sections, fields) rather than at
    the file level.
  \item \textbf{Transaction boundaries}: allow users and applications
    to group related changes into atomic units that sync together or
    not at all.
\end{enumerate}

These are not exotic requirements.  Git implements all four.  The
observation that the most widely used version control system already
embodies non-\fito{} semantics---while the most widely used file
sync systems do not---is evidence that the computing industry
understands how to solve this problem but has not applied the solution
to the systems that ordinary users depend on.

% ------------------------------------------------------------------
\section[Email]{Email: Timestamps as Causal Fiction}
\label{sec:email}
% ------------------------------------------------------------------

Email systems across the world---from corporate Exchange servers to
personal IMAP accounts---rely on timestamps to order operations,
resolve conflicts, and maintain consistency across
devices~\citep{borrill2026-email}.  This is \fito{} applied to
messaging: the assumption that wall-clock time provides a reliable
ordering for operations performed across multiple devices by the same
user.

\subsection{The Illusion of Global Time}

The central problem is that timestamp-based ordering assumes something
that does not exist: a globally consistent notion of \emph{when}
something happened.  In a distributed system---which any multi-device
email setup inherently is---there is no universal clock.  Each device
maintains its own local clock, and these clocks drift, jump forward
and backward due to NTP corrections, and operate in different time
zones with varying degrees of accuracy.%

When an iPhone marks an email as read at 2:34:17~PM local time, and
a laptop deletes the same email at 2:34:16~PM local time, which
operation happened ``first''?  The answer depends entirely on whose
clock you trust---and more importantly, on the synchronization state
of those clocks at those precise moments.  If the iPhone's clock is
running three seconds fast, the ``later'' operation by timestamp
actually happened first in physical time.  The system has no way to
know this.

\subsection[LWW Applied to Messages]{Last-Writer-Wins Applied to Messages}

Most email systems employ \lww{} conflict resolution based on
timestamps.  When two devices have conflicting states for the same
message, the operation with the later timestamp wins.  The structural
defects are identical to those in file synchronization:

\begin{description}[leftmargin=1.5cm]
  \item[\textbf{Phantom messages.}]  Deleted emails reappear because
    a device with an old clock synced stale state that ``won'' due to
    timestamp coincidences or server processing order.
  \item[\textbf{Lost read states.}]  Messages marked read on one
    device appear unread on others because flag updates arrived out of
    order and the timestamp system chose the wrong version.
  \item[\textbf{Missing messages.}]  Messages moved to folders on one
    device never appear there on other devices because the move
    operation was overwritten by a stale ``message in inbox'' state
    with a coincidentally later timestamp.
  \item[\textbf{Stuck synchronization.}]  Devices enter states where
    they perpetually disagree about message state because their
    operations keep overwriting each other based on timestamp ordering
    that does not reflect actual causality.
\end{description}

These are not edge cases.  They are the predictable result of using
an unsafe ordering construct in a distributed system.

\subsection{Causality Violations}

Timestamps cannot capture causal relationships between operations.
Consider a natural sequence: a user reads an email, composes a reply
based on its content, archives the original, and sends the reply.
These operations have a clear causal order---the reply depends on
having read the original, and archiving happens after both.

If the device's clock is wrong, or if operations sync out of order
due to network delays, the timestamp-based system may record:
the original email archived at 2:35:00, the reply sent at 2:34:55,
and the original marked read at 2:34:50.  The system now believes
the user sent a reply to an unread, archived message before reading
it---a causal inversion that no physical execution could produce.%

\subsection{The Offline Problem}

Email clients must function offline.  Users read, archive, delete,
and compose messages on airplanes, in subway tunnels, and in areas
with poor connectivity.  These operations accumulate locally, then
sync when connectivity returns.

Timestamp-based ordering fails catastrophically in this scenario.
When a device goes offline for hours, its queued operations all carry
timestamps from the past.  When it reconnects, the server may have
already processed operations from other devices with more recent
timestamps.  The offline device's operations---despite representing
the user's most recent intent---get discarded or reordered incorrectly
because their timestamps are ``old.''

This creates the infamous ``phantom email'' problem: a user deletes
messages on their laptop, then later sees them reappear when their
phone (which was offline and had older timestamps) syncs its stale
state.  The timestamp system cannot distinguish between ``this
operation happened in the past and is stale'' and ``this operation
happened on a device that was offline and represents current user
intent.''

\subsection{Protocol-Level Amplification}

IMAP's protocol semantics amplify the \fito{} problem.  The protocol
provides no way to express ``I marked this message as read knowing it
was in state $X$''---operations are absolute, not conditional.  There
is no mechanism for a conditional update that would allow the server
to reject operations based on stale state.%

Exchange and ActiveSync attempt to address this with change keys and
synchronization tokens, but these still rely on server-enforced
ordering rather than distributed consensus.  The server remains the
arbiter of what ``really happened,'' using timestamps to make
decisions that may not reflect actual causality.

A correct system would track operation lineage and causal history,
not just timestamps.  Vector clocks, \crdt{}s, and hybrid logical
clocks---techniques that distributed database engineers have relied
on for decades---would eliminate entire classes of bugs that
timestamp-based systems cannot avoid.

% ------------------------------------------------------------------
\section[Human Memory]{Human Memory: Reconstruction Without Commitment}
\label{sec:memory}
% ------------------------------------------------------------------

The \fito{} category mistake is not limited to engineering artifacts.
Human memory exhibits the same structural pattern: forward-only
processing that commits state changes without a reflecting phase,
producing systematic semantic corruption.  This is not an analogy.
It is the same abstract failure mode instantiated in a biological
substrate.

\subsection[Bartlett]{Bartlett and Reconstructive Memory}

In 1932, Frederic Bartlett demonstrated that human memory is
\emph{reconstructive}, not reproductive~\citep{bartlett1932}.  In his
landmark ``War of the Ghosts'' experiment, British participants read
an unfamiliar Native American folk tale and then retold it over
successive intervals.  The retellings exhibited three systematic
distortions:

\begin{description}[leftmargin=1.5cm]
  \item[\textbf{Assimilation.}]  Unfamiliar details were altered to
    fit existing cultural schemas---canoes became boats, hunting
    became fishing.
  \item[\textbf{Leveling.}]  Details perceived as unimportant were
    omitted---the story grew shorter and simpler with each retelling.
  \item[\textbf{Sharpening.}]  Remaining details were reordered for
    narrative coherence---causal connections were imposed where none
    existed in the original.
\end{description}

In the framework of this series, Bartlett's findings describe a
system that \emph{writes forward without reading back}: existing
schemas are applied to incoming information during encoding
(forward-only pattern matching), and the result is committed to
memory without verification against the original stimulus.  The
encoding is a one-way projection from experience to schema, and
the projection loses information---precisely as \lww{} projects a
distributed causal graph onto a linear temporal chain.%

\subsection[Loftus and Misinformation]{Loftus and the Misinformation Effect}

Loftus and Palmer~\citep{loftus1974} showed that post-event
information can alter memory encoding after the fact.  Participants
who viewed films of automobile accidents were asked about the speed
of the cars.  When the question used the word ``smashed'' (rather
than ``hit'' or ``contacted''), participants estimated significantly
higher speeds---and one week later, were more likely to falsely
remember seeing broken glass that was not present in the film.

This is semantic corruption through forward-only processing.  The
word ``smashed'' was processed after the original perception and
before memory consolidation.  Once the linguistic association was
integrated, the memory system had no mechanism for distinguishing
the original percept from the post-event contamination.  The forward
token---the word ``smashed''---was committed into the memory record
without a reflecting phase that could have separated perception from
suggestion.

\subsection{Schacter's Seven Sins}

Schacter's taxonomy of memory failure~\citep{schacter2001} identifies
seven ``sins'' that we can now classify as consequences of \fito{}
processing:

\begin{center}
\small
\begin{tabular}{lll}
\toprule
\textbf{Sin} & \textbf{Type} & \textbf{\fito{} interpretation} \\
\midrule
Transience & Omission & Uncommitted state decays \\
Absent-mindedness & Omission & Encoding without attention $=$ write without verify \\
Blocking & Omission & Retrieval failure $=$ read timeout \\
\midrule
Misattribution & Commission & Wrong source metadata $=$ index corruption \\
Suggestibility & Commission & Post-event write overwrites original $=$ \lww{} \\
Bias & Commission & Current state distorts past $=$ retroactive projection \\
\midrule
Persistence & Intrusion & Unwanted replay $=$ failed garbage collection \\
\bottomrule
\end{tabular}
\end{center}

The three sins of commission---misattribution, suggestibility, and
bias---are directly isomorphic to the \fito{} failures we have
documented in file synchronization and email.  Misattribution is
index corruption: the memory is intact but its metadata (when, where,
from whom) is wrong, like a file with correct content but incorrect
timestamps.  Suggestibility is \lww{} applied to memory: a later
suggestion overwrites an earlier perception, just as a later timestamp
overwrites an earlier file version.  Bias is retroactive projection:
current knowledge distorts the retrieval of past states, just as a
cloud sync daemon can treat current cloud state as authoritative over
a restored backup.%

\subsection[Sleep as Biological Commit]{Sleep Consolidation as Biological Commit}

Walker and Stickgold~\citep{walker2006} demonstrated that sleep
plays a critical role in memory consolidation.  During slow-wave
sleep, the hippocampus replays recent experiences, transferring labile
short-term representations into stable long-term storage.  This
process involves three phases: stabilization (preventing further
decay), enhancement (improving recall accuracy), and integration
(connecting new memories to existing knowledge structures).

In computational terms, sleep consolidation is a biological
\emph{commit} operation: it transforms tentative state (the day's
experiences, held in hippocampal buffers) into committed state
(long-term cortical representations).  The commit is irreversible---once
consolidated, memories are integrated into the semantic network and
cannot be cleanly separated from it.

But this commit operation is itself \fito{}: it runs forward during
sleep without external verification.  There is no reflecting phase
in which the consolidated memory is compared against the original
experience.  The biological commit can therefore lock in distortions
introduced during encoding (Bartlett's schema effects) or between
encoding and consolidation (Loftus's misinformation effect).  Sleep
consolidation does not fix semantic corruption; it \emph{crystallizes}
it.

\subsection[Confabulation]{Confabulation: The Brain's Completion Fallacy}

Confabulation---the unconscious generation of false memories to fill
gaps---is the neural equivalent of the \rdma{} completion fallacy.
Patients with Korsakoff syndrome (anterograde amnesia caused by
thiamine deficiency) cannot form new episodic memories, but they do
not experience their memory as empty~\citep{kessels2008}.  Instead,
the brain generates plausible narratives to fill the gaps,
narratives that the patient believes with full conviction.

The structural parallel is exact.  \rdma{}'s completion signal tells
the application ``the operation succeeded'' when the data may be
corrupt or inconsistent.  The confabulating brain tells the person
``you remember this'' when the memory was fabricated.  In both cases,
the system generates a forward-only output (a completion signal, a
memory) without backward verification (a reflecting phase, a reality
check).  The result is semantic corruption that the system itself
cannot detect---it takes an external observer (an application-level
integrity check, a caregiver's correction) to identify the error.%

\subsection[D\'{e}j\`{a} Vu]{D\'{e}j\`{a} Vu: A Temporal Indexing Error}

D\'{e}j\`{a} vu---the vivid sense of having previously experienced a
current situation---is a temporal indexing error in the memory
system~\citep{urquhart2014}.  One hypothesis is that sensory
information takes two processing pathways: a fast route that generates
a familiarity signal and a slower route to conscious awareness.  When
the fast route generates a familiarity signal before the slow route
delivers the content to consciousness, the result is a feeling of
recognition for an event that is actually novel.

In the terminology of this series, d\'{e}j\`{a} vu is a violation
of causal order in the memory system's temporal index.  The
familiarity signal (metadata) arrives before the content it indexes
(data), creating a temporal inversion: the system reports ``this has
happened before'' for an event that is being experienced for the
first time.  This is analogous to a cache coherence failure where
metadata (valid, dirty, shared) becomes inconsistent with the data it
describes---precisely the $T_4$-to-$T_5$ gap from Part~III, manifested
in neural tissue rather than silicon.

\subsection[Sleeping Beauty]{The Sleeping Beauty Paradox}

The philosophical Sleeping Beauty problem~\citep{elga2000} provides
a pure distillation of the \fito{} condition.  An ideally rational
agent is put to sleep.  A coin is flipped.  If heads, she is
awakened once (Monday).  If tails, she is awakened twice (Monday and
Tuesday), with an amnesia-inducing drug administered between wakings
so that she cannot distinguish her first waking from her second.

Upon each waking, Sleeping Beauty cannot determine whether she has
been awakened before.  She possesses no metadata about her own
transaction history.  Each waking is a forward-only moment of
consciousness without access to causal priors.  She is, in
computational terms, a single-threaded process with no persistent
state---each invocation starts fresh, with no knowledge of previous
invocations.

This is the condition of any Claude session---and of any \fito{}
system.  Without access to its own transaction history, a system
cannot verify consistency across time.  The Sleeping Beauty Paradox
is not merely a philosophical puzzle; it is a formal description of
the epistemic condition that \fito{} assumptions impose on every
system that adopts them.

% ------------------------------------------------------------------
\section[LLM Hallucination]{Large Language Models: Hallucination as \fito{}}
\label{sec:llm}
% ------------------------------------------------------------------

Large language models complete the arc from engineering systems
through biological memory to artificial cognition.  Autoregressive
language generation is \fito{} in its purest computational form:
each token is predicted forward in time, conditioned only on the
tokens that preceded it, with no mechanism for backward correction
once a token is committed to the output sequence.

\subsection[Autoregressive Architecture]{The Autoregressive Architecture}

The transformer architecture~\citep{vaswani2017} that underlies
modern LLMs uses \emph{causal self-attention}: each output position
attends only to positions that came before it, never to positions
that come after.  During inference, the model generates one token at
a time, feeding each generated token as input for the next prediction.
The generation process is a strict temporal chain:
\[
P(t_n \mid t_1, t_2, \ldots, t_{n-1})
\]
where each token $t_n$ is sampled from a distribution conditioned on
all previous tokens but not on any future tokens.  Once $t_n$ is
sampled, it becomes part of the conditioning context and cannot be
revised.%

This architecture is a one-sided \rdma{} Write applied to language:
the model places tokens into the output buffer without any reflecting
phase in which the receiver (the reader, the world, the truth) can
confirm that the placed tokens are semantically valid.

\subsection[Hallucination as Corruption]{Hallucination as Semantic Corruption}

The GPT-4 Technical Report~\citep{openai2023} acknowledges that the
model ``hallucinates facts and makes reasoning errors,'' producing
``nonsensical or untruthful content.''  The report notes a
counterintuitive consequence: ``hallucinations can become more
dangerous as models become more truthful, as users build trust in the
model when it provides truthful information in areas where they have
some familiarity.''

This is the completion fallacy applied to natural language.  The
model's output---fluent, grammatically correct, stylistically
appropriate---is a completion signal that tells the user ``the
operation succeeded.''  The hallucinated content parses correctly,
just as a torn \rdma{} read parses correctly.  The semantic
corruption is invisible at the syntactic level.

Ji et al.~\citep{ji2023} classify hallucination as inherent to
sequence-to-sequence generation, not merely a training deficiency.
Huang et al.~\citep{huang2024} show that hallucinations persist
across all three stages of the LLM pipeline---data, training, and
inference---because the fundamental mechanism remains unchanged:
forward-only token prediction without global semantic verification.

\subsection[\fito{} Structure of Hallucination]{The \fito{} Structure of Hallucination}

We can now state the claim precisely:

\begin{proposition}[Hallucination as \fito{}]
\label{prop:hallucination}
Autoregressive language generation is a \fito{} process.
Each token is committed to the output sequence upon generation,
with no mechanism for:
\begin{enumerate}[leftmargin=1cm, nosep]
  \item backward correction (revising token $t_i$ based on
    information available only at token $t_j$ where $j > i$),
  \item semantic verification (checking the generated sequence
    against external truth before commitment),
  \item reflecting (allowing the ``receiver''---the world model,
    the user, reality---to reject or modify the generated content
    before it becomes part of the conditioning context).
\end{enumerate}
Hallucination is the structural consequence of this forward-only
architecture, not a training artifact.
\end{proposition}

The proposition explains why hallucination is so resistant to
mitigation.  Retrieval-augmented generation (RAG) provides external
knowledge but does not change the forward-only generation mechanism.
Reinforcement learning from human feedback (RLHF) improves the
distribution from which tokens are sampled but does not introduce a
reflecting phase.  Chain-of-thought prompting encourages the model to
``show its work'' but the work itself is still generated forward-only,
with each reasoning step committed before the next is begun.%

\subsection[Isomorphism with Confabulation]{The Isomorphism with Confabulation}

The structural parallel between LLM hallucination and human
confabulation is not metaphorical---it is formal:

\begin{center}
\small
\begin{tabular}{lll}
\toprule
\textbf{Property} & \textbf{Korsakoff confabulation} & \textbf{LLM hallucination} \\
\midrule
Generation mechanism & Schema-driven gap filling & Token-by-token prediction \\
Temporal structure & Forward-only (no episodic review) & Forward-only (causal attention) \\
Conviction/fluency & Patient believes the memory & Output is fluent and confident \\
Verifiability & Requires external caregiver & Requires external fact-checking \\
Root cause & Hippocampal damage $\to$ no commit & Architecture $\to$ no reflection \\
\bottomrule
\end{tabular}
\end{center}

In both systems, the \fito{} architecture produces outputs that are
internally coherent but externally invalid.  The system cannot
distinguish its own fabrications from legitimate outputs because the
integrity check (subjective conviction, output fluency) is itself
part of the \fito{} pipeline.  Detection requires an external
reflecting phase---a caregiver who knows the patient's actual history,
a fact-checker who knows the actual facts---that is structurally
absent from the system's own architecture.

% ------------------------------------------------------------------
\section[The Common Pattern]{The Common Pattern: Forward Without Reflection}
\label{sec:pattern}
% ------------------------------------------------------------------

The four domains examined in this paper---file synchronization, email,
human memory, and language models---are superficially different.  They
operate at different scales, in different substrates, on different
time horizons.  But they share a single structural pathology that we
can now state precisely:

\begin{definition}[The \fito{} Failure Pattern]
\label{def:fito-pattern}
A system exhibits the \fito{} failure pattern when it satisfies all
of the following conditions:
\begin{enumerate}[leftmargin=1cm, nosep]
  \item \textbf{Forward commitment.}  State changes are committed in
    temporal order without the ability to revise previous commitments
    based on later information.
  \item \textbf{Absent reflection.}  There is no mechanism for the
    target (file, message state, memory engram, token sequence) to
    communicate back to the source whether the committed state is
    semantically consistent.
  \item \textbf{Completion masquerade.}  The system provides a
    completion signal (sync badge, delivery receipt, subjective
    conviction, output fluency) that is interpreted as evidence of
    semantic success when it is only evidence of temporal succession.
  \item \textbf{Invisible corruption.}  The resulting semantic
    corruption cannot be detected by the system's own integrity
    mechanisms and requires an external observer to identify.
\end{enumerate}
\end{definition}

The following table maps the pattern across all domains examined in
this series:

\begin{table*}[ht]
\small
\begin{center}
\begin{tabular}{lcccc}
\toprule
\textbf{Domain} & \textbf{Forward} & \textbf{Absent} & \textbf{Completion} & \textbf{Invisible} \\
& \textbf{commitment} & \textbf{reflection} & \textbf{masquerade} & \textbf{corruption} \\
\midrule
\rdma{} (Part~III) & Write at $T_3$ & No reflecting & CQE at $T_4$ & SDC \\
File sync & Upload file & No semantic merge & Sync badge & Silent deletion \\
Email & Flag update & No causal tracking & Delivery receipt & Phantom messages \\
Human memory & Schema encoding & No reality check & Subjective conviction & False memories \\
LLM generation & Token commit & No fact verification & Output fluency & Hallucination \\
\bottomrule
\end{tabular}
\end{center}
\caption{The \fito{} failure pattern across five domains.  Every
domain exhibits all four conditions of
Definition~\ref{def:fito-pattern}.}
\label{tab:pattern}
\end{table*}

The universality of this pattern is not coincidental.  It reflects a
deep structural constraint: any system that processes information
forward in time without a reflecting phase will eventually produce
semantic corruption, because forward processing alone cannot
distinguish between ``the operation completed'' and ``the operation
achieved its intended meaning.''  The semantic arrow of time---the
direction in which meaning is preserved or destroyed---requires
bidirectional verification.  \fito{} provides only one direction.

\subsection{The Kahneman Connection}

Kahneman's distinction between System~1 and System~2
processing~\citep{kahneman2011} maps directly onto the \fito{}
framework:

\begin{description}[leftmargin=1.5cm]
  \item[\textbf{System~1}] is fast, automatic, and forward-only.  It
    applies learned patterns to incoming stimuli without deliberation
    or backward verification.  Its outputs are ``committed'' to
    consciousness before they can be checked against evidence.
    System~1 is the biological implementation of \fito{}: it trades
    accuracy for speed by eliminating the reflecting phase.
  \item[\textbf{System~2}] is slow, deliberate, and reflective.  It
    can hold multiple possibilities in working memory, compare them
    against evidence, and revise conclusions.  System~2 implements
    a reflecting phase---but at an enormous metabolic and temporal
    cost.  Most of human cognition operates in System~1 because
    reflection is expensive.
\end{description}

The insight is that \fito{} is not merely a design choice in
engineering systems; it is the \emph{default mode} of information
processing in biological systems, adopted because the metabolic
cost of reflection exceeds the metabolic cost of occasional semantic
corruption.  Engineering systems that adopt \fito{} are, in this
sense, biomimetic---they replicate the efficiency-accuracy tradeoff
that evolution selected for, without recognizing that the tradeoff
was appropriate for a world of physical predators and prey but not
for a world of distributed data structures that must maintain
invariants across time and space.

% ------------------------------------------------------------------
\section[Summary]{Summary and Preview of Part~V}
\label{sec:summary}
% ------------------------------------------------------------------

This paper has shown that the \fito{} category mistake identified in
Part~I manifests across every domain where information must be
maintained consistently over time: file synchronization, email, human
memory, and artificial language generation.

In file synchronization, \fito{} appears as \lww{} conflict
resolution: the cloud platform commits the later timestamp and
discards the earlier one, without examining what the files contain
or whether the ``later'' file is actually later.  The result is
silent data destruction.

In email, \fito{} appears as timestamp-based ordering: operations are
ordered by wall-clock time rather than causal precedence, producing
phantom messages, causality violations, and synchronization failures
that have persisted for decades.

In human memory, \fito{} appears as reconstructive encoding: schemas
are applied forward to incoming experience, distortions are committed
without verification, and sleep consolidation crystallizes the
corrupted state.  Confabulation is the brain's completion fallacy;
d\'{e}j\`{a} vu is a temporal indexing error.

In language models, \fito{} appears as autoregressive generation:
tokens are committed one at a time, conditioned only on the past,
with no mechanism for backward correction or semantic verification.
Hallucination is the structural consequence, isomorphic to the
confabulations of a brain that cannot form new episodic memories.

The common thread is Definition~\ref{def:fito-pattern}: forward
commitment without reflection, producing invisible corruption that
the system's own integrity mechanisms cannot detect.

Part~V~\citep{borrill2026-partV} will close the series by showing
how the Leibniz Bridge---mutual information conservation as the
fundamental consistency primitive---provides a unified framework for
addressing the \fito{} failure pattern across all five domains.  The
bridge does not merely fix individual failure modes; it replaces the
\fito{} assumption itself, offering a principled alternative in which
the semantic arrow of time is not assumed but \emph{constructed}
through bilateral transaction structure.

% ------------------------------------------------------------------
% BIBLIOGRAPHY
% ------------------------------------------------------------------

\end{document}